\documentclass[conference]{IEEEtran}
\IEEEoverridecommandlockouts
\usepackage{cite}
\usepackage{amsmath,amssymb,amsfonts}
\usepackage{algorithmic}
\usepackage{graphicx}
\usepackage{textcomp}
\usepackage{xcolor}
\usepackage{url}
\usepackage{listings}
\usepackage{epigraph}
\usepackage{courier}
\def\BibTeX{{\rm B\kern-.05em{\sc i\kern-.025em b}\kern-.08em
    T\kern-.1667em\lower.7ex\hbox{E}\kern-.125emX}}
\begin{document}

\title{\textit{FaaS and Furious}: abstractions and differential caching for efficient data pre-processing
\thanks{Pre-print of the paper accepted at \textit{DEMAI@IEEE Big Data 2024}.}}

\author{\IEEEauthorblockN{Jacopo Tagliabue}
\IEEEauthorblockA{
\textit{Bauplan Labs}\\
New York, US \\
jacopo.tagliabue@bauplanlabs.com}
\and
\IEEEauthorblockN{Ryan Curtin}
\IEEEauthorblockA{
\textit{Bauplan Labs}\\
Atlanta, US \\}
\and
\IEEEauthorblockN{Ciro Greco}
\IEEEauthorblockA{
\textit{Bauplan Labs}\\
New York, US \\}
}

\maketitle

\begin{abstract}
Data pre-processing pipelines are the bread and butter of any successful AI project. We introduce a novel programming model for pipelines in a data lakehouse, allowing users to interact declaratively with assets in object storage. Motivated by real-world industry usage patterns, we exploit these new abstractions with a columnar and differential cache to maximize iteration speed for data scientists, who spent most of their time in pre-processing -- adding or removing features, restricting or relaxing time windows, wrangling current or older datasets. We show how the new cache works transparently across programming languages, schemas and time windows, and provide preliminary evidence on its efficiency on standard data workloads.\end{abstract}


\section{Introduction}

\setlength{\epigraphrule}{0pt}
\epigraph{Scans do not repeat themselves, but they often rhyme.}{(almost) Mark Twain}

\begin{figure}
\centerline{\includegraphics[width=0.45\textwidth]{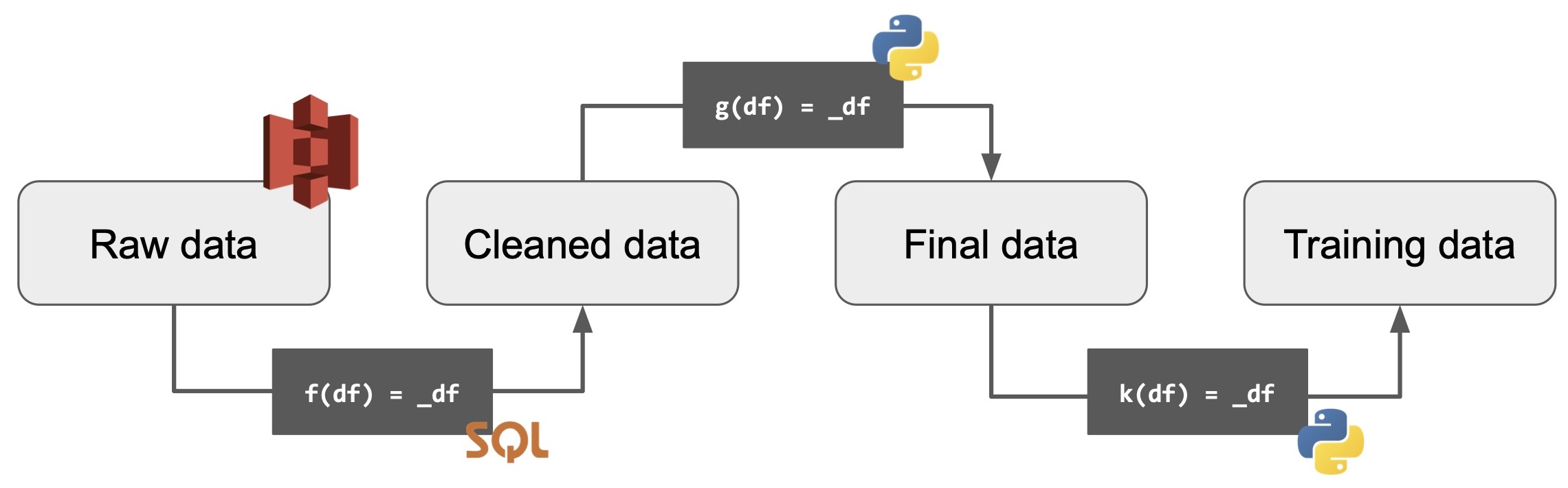}}
\caption{\textbf{A sample \textit{multi-language}, \textit{cloud} data pipeline}. The pipeline takes raw data in object storage (S3) to a final training dataset, by going through intermediate steps that wrangle dataframes into progressively cleaner data assets.}
\label{fig:flow}
\end{figure}

As the already large market for analytics, Business Intelligence and Artificial Intelligence keeps increasing \cite{carganalytics}, the community is once again re-discovering the pivotal role of data preparation pipelines for the success of any data-driven initiative \cite{10.1145/3411764.3445518}. 
In recent years, the lakehouse architecture  \cite{Zaharia2021LakehouseAN}
and, more generally, the decoupling of data and compute became the standard for unified data processing at the enterprise scale: a pre-processing pipeline takes the shape of a directed acyclic graph (DAG), in which nodes are \textit{transformations}, i.e. functions from dataframe(s) \cite{10.14778/3407790.3407807} to dataframe which represent the cleaning, aggregation and simplification logic when going from ``raw'' to ``cleaned'' dataframes for downstream consumption, e.g. training a Machine Learning model (Fig.~\ref{fig:flow}).

As a fast feedback loop is generally recognized as the hallmark of successful data initiatives \cite{https://doi.org/10.48550/arxiv.2209.09125, Tagliabue2023ReasonableSM}, it is imperative that data scientists can easily experiment with different languages, libraries, and data subsets. Notwithstanding the industry interest in solving the problem \cite{cargpipeline}, popular pipeline frameworks (e.g. \cite{airflow} \cite{luigi}) are mostly designed for asynchronous machine execution (e.g. batch jobs at night), and not for iterative work, leaving data scientists to work on small samples and then hand off the project for production refactoring, or roll their own productivity abstractions and constantly re-inventing the wheel.

In this short paper, we discuss how to design pipeline abstractions to optimize for user interactivity, and then dive deeper on a columnar cache design that significantly improves performance in iterated workflows. In particular, our main contributions are the following:

\begin{enumerate}
    \item we introduce \texttt{Bauplan} as a pipeline tool, and discuss its syntax and semantics to run DAGs on top of object storage; in particular, we highlight the distinction between dataframes as logical abstractions \textit{vs.} dataframes as physical operations; 
    \item we identify data movement as a primary source of latency for data workloads, and argue that scans over object storage are the atomic building blocks for a heterogeneous set of operations;
    \item we share design choices and preliminary results for our differential cache for cloud tables; our cache works transparently across dataset versions, sets of projections and overlapping filters. Finally, we showcase its efficacy with preliminary quantitative benchmarks: compared to baseline, the proposed design allows the system to read up to 30\% fewer bytes. 
\end{enumerate}

Importantly, as our solution is built with open source lakehouse formats in mind (Apache Parquet \cite{parquet}, \textit{Apache Iceberg} \cite{iceberg})), our insights can be used to improve other cloud-first systems with minimal adaptations.

\section{Pipeline building on the lakehouse}
\label{sec:pipeline}

\textit{Bauplan} is a data lakehouse platform for running queries and declarative pipelines, comprising an Iceberg-compatible data catalog \cite{10.1145/3650203.3663335}, a data-aware Function-as-a-Service (FaaS) runtime \cite{Tagliabue2023BuildingAS}, and a novel set of abstractions for DAGs. Users express data transformations as Python functions (or SQL queries) with the signature $f(dataframe(s)) \rightarrow dataframe$: for example, a pure Python implementation of the DAG in Fig.~\ref{fig:flow} would look like the following:

\addvspace{\baselineskip}
\begin{lstlisting}[showstringspaces=false,columns=fullflexible,language=Python,caption=A sample DAG in Bauplan]
@bauplan.model()
@bauplan.python("3.10", pip={"pandas": "2.0"})
def cleaned_data(
    # reference to its parent DAG node
    data=bauplan.Model(
        "raw_data",
        columns=["c1, "c2", "c3"], 
        filter="eventTime BETWEEN 2023-01-01 AND 2023-02-01"
    ) 
):
   # transformation logic goes here...
   return data.do_something()

@bauplan.model()
@bauplan.python("3.11", pip={"pandas": "2.0"})
def final_data(
    data=bauplan.Model("cleaned_data") 
):
   return data.do_something()

@bauplan.model()
@bauplan.python("3.11", pip={"pandas": "1.0"})
def training_data(
    data=bauplan.Model("final_data") 
):
    return data.do_something()
\end{lstlisting}
\addvspace{\baselineskip}

\begin{figure}
\centerline{\includegraphics[width=0.45\textwidth]{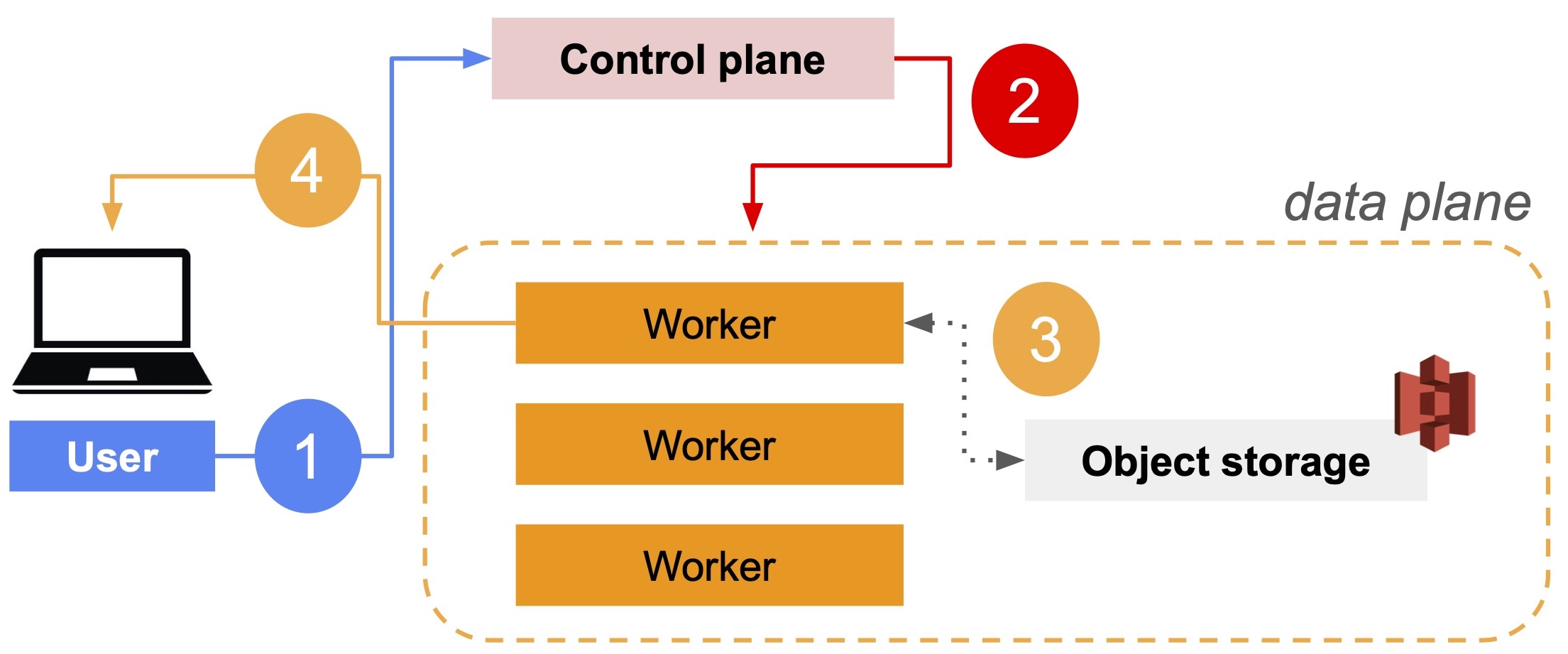}}
\caption{\textbf{High-level communication flow between users and the cloud platform}. 1) user requests a DAG execution, 2) the control plain sends a physical plan to a cloud worker, 3) the worker fetches data from object storage and 4) returns the log messages and the tuples back to the user.}
\label{fig:bauplan}
\end{figure}

The DAG structure is \textit{implicitly} expressed through function inputs, and reconstructed by the platform when the code is submitted. As for the relevant runtime properties, a decorator allows to express the desired Python interpreter and dependencies. While the runtime details are beyond the scope of this paper, it is important to have a high-level mental model of the cloud context where the cache will be placed. Similarly to a cloud database (and unlike existing FaaS platforms), \textit{Bauplan} pipelines happen across a multi-tenant \textit{control plane} and a single-tenant secure \textit{data plane} (Fig.~\ref{fig:bauplan}): user code is parsed by the control plane, which produces a \textit{physical plan} with the required low-level operations. Finally, workers in the customer data plane execute the plan and return results to the user. The declarative nature of the pipeline accomplishes two goals: first, it creates a principled division of labor between the system (infrastructure and optimization) and the data scientist (business logic and choice of language / libraries). Second, it is a necessary decoupling to run the same pipeline over different versions of the same table (e.g. running today's code on last Friday's rows \cite{10.1145/3650203.3663335}), or different \textit{physical realizations} of the same asset (reading data from S3 or the cache, as described below). 

Before examining a sample physical plan, it is worth comparing this programming model with non-data-aware frameworks, which couple the physical representation with code, as shown for example in the Airflow reference implementation for pre-processing by \textit{AWS} \cite{awsairflow}:

\addvspace{\baselineskip}
\begin{lstlisting}[showstringspaces=false,columns=fullflexible,language=Python,caption=Simplified snippet for pre-processing in an Airflow DAG.]

def preprocess(
       s3_in_url, s3_out_bucket, s3_out_prefix
    ):
    # Do pre-processing and save the result in
    # "s3_out_bucket / s3_out_prefix"
    return "SUCCESS"

preprocess_task = PythonOperator(
    task_id="preprocessing",
    dag=dag,
    python_callable=preprocess.preprocess
    )
\end{lstlisting}
\addvspace{\baselineskip}

Not only the pre-processing function operates at the physical level of S3 files (instead of the logical level of dataframes), but it saves its output to \textit{s3\_out\_bucket} as a side effect, instead of returning the cleaned dataframe to the caller, preventing any further optimization.

\subsection{Physical plan and object storage}

\begin{figure}
\centerline{\includegraphics[width=0.47\textwidth]{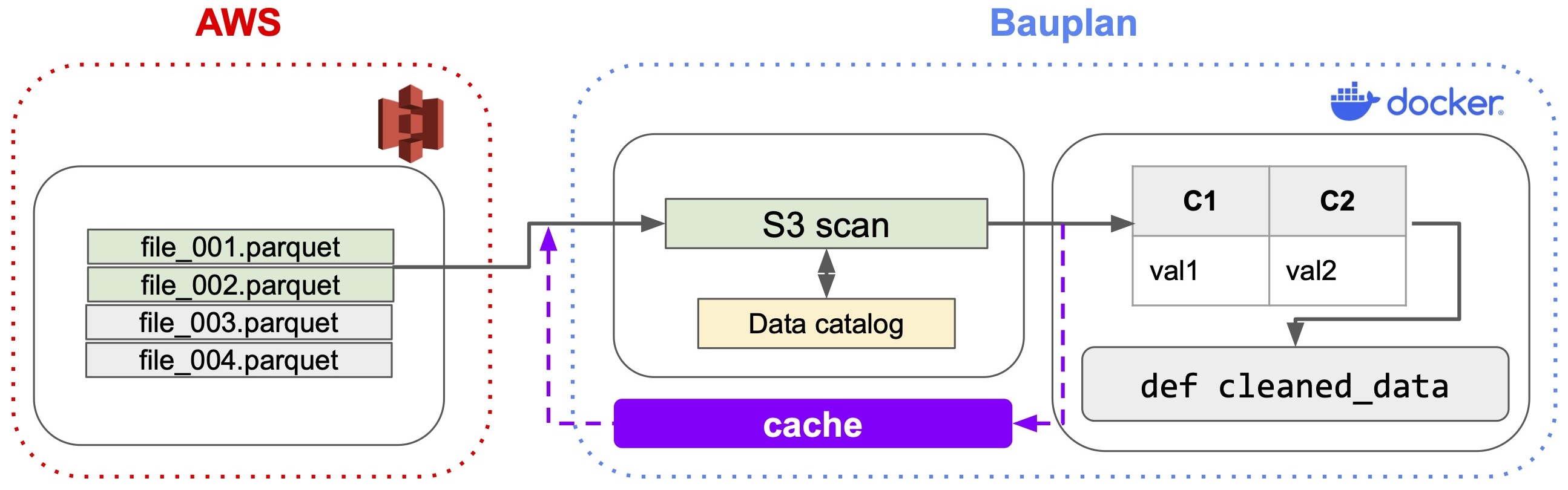}}
\caption{\textbf{The \textit{physical plan} for \textit{cleaned\_data}.}. A system function performing scans over S3 is added automatically \textit{before} the user code: this decoupling shields users from data management and allows the addition of a data cache (purple, Section~\ref{sec:design}).}
\label{fig:physical}
\end{figure}

The declarative APIs leave, by design, a gap between the dataframes as logically expressed in the code, and machine instructions on how to fetch them from object storage. The easiest way to understand why caching is fundamental is to inspect a simple plan for the \textit{cleaned\_data} step (Fig.~\ref{fig:physical}), by working backwards from the user code to the required physical assets:

\begin{enumerate}
    \item user code in the \textit{cleaned\_data} function receives as input the variable \textit{data} as a two column Arrow table \cite{arrow} representing a scan over  \textit{raw\_data};
    \item the table is provided by a \textit{system} function -- i.e. a \textit{Bauplan} function that gets executed on behalf of the user behind the scenes: this function translates the request for \textit{raw\_data} into a set of projections and filters over Parquet files, and finally performs the range-byte reads from S3 to get the uncompressed dataframe.
\end{enumerate}

The separation between logical and physical representation of inputs frees users from dealing with complex engineering tasks (efficient S3 readings and data passing) and opens the door for system-level optimizations when the same physical representation can be re-used across runs. As data volume grows, the bottleneck clearly becomes the time to read and decompress data from files in S3 (Table~\ref{tab1:reads}): no other optimization matters if we can't lower the latency of moving data to user functions. In other words, to achieve the fast turn-around data scientists need, we need to turn our attention to \textit{caching}.

\section{Differential Caching}
\label{sec:caching}

\subsection{Design principles}
\label{sec:design}

The two major choices in designing a data cache are which objects should be cached and what their physical representation should be. A straightforward strategy would be to memoize the pair \textit{inputs + output dataframe} (a so-called ``result cache'' in the database community \cite{10.1145/3626246.3653395}), and store Parquet files in local SSDs; however, as we shall see, much better choices can be made in this context.

\textbf{Scans as primary objects}. Surveys and traces from both the Analytics and Machine Learning communities \cite{Xin2018HowDI} stress the prevalence of repeatedly query the same data assets when doing data preparation \cite{Renen2024}. However, consider these three workloads in a sequence:

\begin{enumerate}
    \item \textit{user A} runs our Python DAG above, with \texttt{c1, c2, c3} as projections and a date filter \texttt{eventTime BETWEEN 2023-01-01 AND 2023-02-01};
    \item \textit{user B} runs the SQL query (effectively, a one-node DAG): \texttt{SELECT c1, \textbf{c3} FROM raw\_data WHERE eventTime BETWEEN 2023-01-01 AND 2023-\textbf{03}-01};
    \item \textit{user A}, unconvinced by her previous feature set, re-runs the DAG with \texttt{\textbf{c2}} as the only projection and a day-wide filter for debugging, \texttt{eventTime BETWEEN 2023-01-01 AND 2023-01-02}.
\end{enumerate}

\begin{figure}
\centerline{\includegraphics[width=0.48\textwidth]{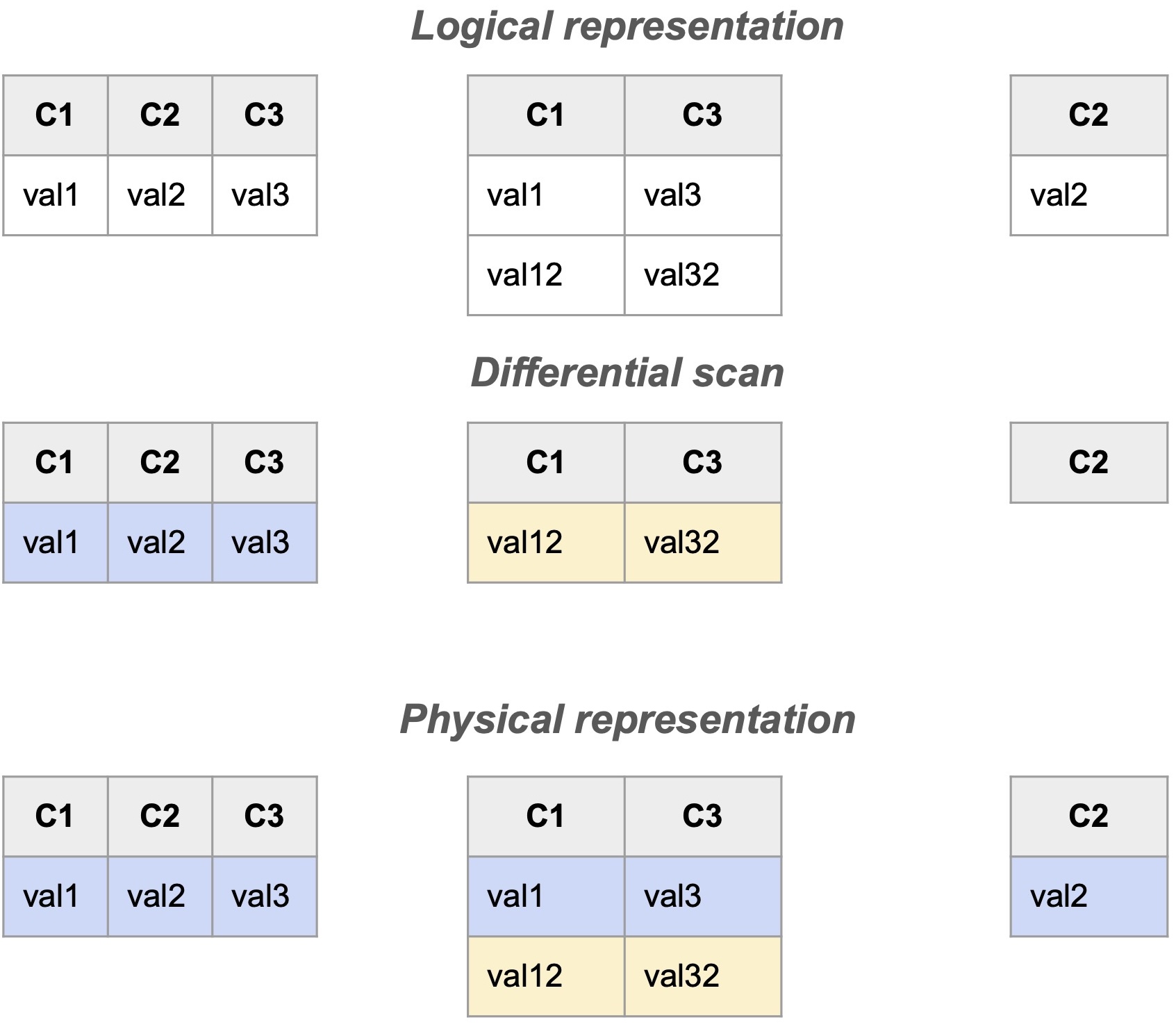}}
\caption{\textbf{Differential, language-agnostic scans for workloads (1)-(3) (left to right).}: logical representation of the dataframes based on user code (\textit{top}); S3 scans to download dataframe fragments (\textit{middle}, request \#3 requires no scan); physical dataframes as assembled from fragments (\textit{bottom}).}
\label{fig:scans}
\end{figure}

If the cache operates on \textit{input + tuples}, those operations would all trigger novel scans over object storage, as all inputs are different: the iterative nature of data pre-processing within a team is such that scans often do not exactly repeat, but they still have enough in common to invite re-use, across \textit{both users and languages}. Assuming an efficient way to identify scans and a representation for dataframes as built from data fragments, the ideal execution plan is depicted in Fig.~\ref{fig:scans}: for \textit{user B}, the system retrieves the missing dates for an existing set of projections (i.e., \texttt{February 2023}), and for the last run no reads from object storage are necessary at all, since the data to fulfill \textit{A}'s request is already in the cache.

\textbf{Arrow table as physical representation}. As scans are performed over object storage as range-byte requests over Parquet files (Fig.~\ref{fig:physical}), storing them in the cache as is seems a natural choice. However, we opted to use Arrow as our physical representation for our cache for two main reasons. \textit{First}, our ideal physical format should allow ``views'' over pre-scanned fragments (Fig.~\ref{fig:scans}), such that downstream functions can zero-copy access their input dataframes: Arrow zero-copy sharing is not just efficient in terms of latency, but avoids expensive memory operations when multiple user functions needs to read from the same scan, as the same Arrow view is shared (or memory-mapped, in case of disk support) by \textit{k} different children.

\textit{Second}, since Arrow is the in-flight representations for dataframes fed to user code, an Arrow-backed cache avoids incurring the decompression and serialization costs of the Parquet-to-Arrow conversion twice -- once during the first read from S3, the second when reading from the cache again. Table~\ref{tab1:reads} reports benchmarks on storage and serialization options when moving data at rest into data in-process (i.e. supplying the input Arrow table to user functions)\footnote{Benchmarking code is available at: \url{https://gist.github.com/jacopotagliabue/57bb14c675a5375338d4a57a88cea32a}.}: it is easy to realize both that IPC (if possible) is incredibly efficient, and that serialization cost is significant even in the absence of the (also significant) latency from S3. Taken together, these two observations vindicate the choice of Arrow from a performance standpoint, on top of purely functional considerations.   


\begin{table}
\caption{Moving dataframes into a user function (\textit{c5.9xlarge})}
\begin{center}
\begin{tabular}{|c|c|c|}
\hline
\textbf{\textit{Rows (Arrow size)}} & \textbf{\textit{Source}}& \textbf{\textit{Seconds (SD)}} \\
\hline
10M (6 GB) & Parquet file in S3& 1.26 (0.14) \\
 & Parquet file on SSD& 0.92 (0.09) \\
 &  Arrow Flight & 0.96 (0.01) \\
 &  Arrow IPC & \textbf{0.0 (0.0)} \\
\hline
50M (30 GB) & Parquet file in S3& 6.14 (0.98) \\
 & Parquet file on SSD& 4.37 (0.15) \\
 &  Arrow Flight & 4.69 (0.01) \\
 &  Arrow IPC & \textbf{0.03 (0.01)} \\
\hline
\end{tabular}
\label{tab1:reads}
\end{center}
\end{table}

\subsection{Implementation}

Based on our two principles -- S3 scans as building blocks, and Arrow tables as physical data representation -- we now describe the implementation of our solution. System-wise, the additional module is trivially integrated into the existing runtime by providing optional reads / writes to the system function (Fig.~\ref{fig:physical}) -- thanks to the programming model (Section~\ref{sec:pipeline}), no re-factoring is needed in any user DAG.

Internally, user code is parsed to recognize scans to optimize -- i.e. as input to \textit{cleaned\_data}, we need an S3 scan over the \textit{raw\_data} Iceberg table, with projections ({\tt \small c1, c2, c3}), and filters ({\tt \small WHERE eventTime BETWEEN 2023-01-01 AND 2023-02-01}). For a given scan, the cache is first filtered to contain only cache elements with matching namespace and table, and whose projections are a superset of the scan's projections. Then, cache elements are applied in a greedy manner (see the pseudo-code below). In the best case, cache elements fully cover the scan to be performed, and no communication with S3 is necessary. Otherwise, the parts of the scan that are not covered by cache elements are requested from object storage, and the results are stored as a new cache element. Cache elements with overlapping or adjacent filters can then be combined.  Importantly, since scans over Iceberg tables are mapped to an underlying set of \textit{immutable} Parquet files, cache invalidation is `free': if a table is modified, the cache can deterministically detect which old entries are not relevant anymore by simply storing pointers to original S3 files, together with scan inputs. \textit{Bauplan}'s native support for Iceberg allows the cache to act both at the logical (leveraging the relation algebra semantics implicit in projections and filters) and the physical level (leveraging file immutability to detect data staleness).

\addvspace{\baselineskip}
\begin{lstlisting}[showstringspaces=false,columns=fullflexible,language=Python,caption=Python pseudo-code for the cache.]

# We assume that `cache` has been pre-filtered to
# contain only relevant elements.
def apply_cache(scan_filter, cache):
    best_filter = None
    best_cost = Inf
    best_cache_elem = None
    # Greedily find the cache element that reduces
    # the cost of the scan the most.  'compute_cost()'
    # returns either the size of the required scan or
    # a bound on the size.
    for e in cache:
        new_filter, new_cost = compute_cost(
            f'({scan_filter}) AND NOT ({e.filter})')

        if new_cost < best_cost:
            best_filter = new_filter
            best_cost = new_cost
            best_cache_elem = e

    # Recursively apply more filters, unless the scan
    # filter is 'WHERE FALSE', meaning that no scan
    # is needed.
    if best_filter == 'FALSE':
        return best_filter, best_cost # cost is 0.
    elif best_filter != None:
        # In this case we applied a cache element but
        # a scan is still needed. 
        return apply_cache(best_filter,
            cache.remove(best_cache_elem))
\end{lstlisting}
\addvspace{\baselineskip}

Although a non-greedy algorithm will produce the same scan after the cache, it is better to use a greedy algorithm
as using fewer cache elements will result in a smaller {\tt \small UNION} operation to combine them together. As a working example, consider the workflow from Section~\ref{sec:design}: after A's first run, the scan is stored in the cache as {\tt \small cache\_entry\_1}; when B runs her workload, {\tt \small cache\_entry\_1} will be re-used automatically: the system builds the relevant Arrow table by combining the cached entry with the re-written S3 scan: 
\begin{lstlisting}[showstringspaces=false,columns=fullflexible,language=sql]
SELECT c1, c3 FROM
  (SELECT c1, c3 FROM cache_entry_1 UNION
   SELECT c1, c3 FROM t WHERE eventTime BETWEEN 2023-02-02 AND 2023-03-01);
\end{lstlisting}

\subsection{Preliminary benchmarks}
As we cannot publish private customer traces, we report preliminary results on both the TPC-H 22 queries (scale factor 1 and 100), as well as our motivating scenario from Section~\ref{sec:design}, which we operationalize on real data using the NYC Taxi Dataset for 2023\footnote{\url{https://www.nyc.gov/site/tlc/about/tlc-trip-record-data.page}}, with \textit{projections=(hvfhs\_license\_num, PULocationID, DOLocationID)} and datetime filtering on  \textit{pickup\_datetime}. We compare our differential cache against
a result cache -- which caches tuples under the hash of the exact input parameters --, and a scan cache -- which caches the results of S3 scans exactly (which may or may not be equal to the fully specified input parameters). Both result and scan caches are common in real-world data systems, such as cloud warehouses \cite{10.1145/3626246.3653395}.

%

\begin{table}
\begin{center}
\begin{tabular}{|c|c|c|c|}
\hline
{\bf Workload} & {\bf Result cache} & {\bf Scan cache} & {\bf Bauplan }  \\
\hline
TPC-H, SF1                               & 2.722 & 2.719 & \textbf{2.252} (17.1\%) \\
TPC-H, SF100                             & 323.5 & 323.3 & \textbf{257.7} (20.2\%)\\
Sec.~\ref{sec:design} workload & 1.703 & 1.703 & \textbf{1.171} (31.2\%) \\
\hline
\end{tabular}
\end{center}
\caption{Comparing result cache, scan cache and \textit{Bauplan}'s differential cache based on total GBs processed (lower is better - percentages show the savings in bytes vs scan cache).}
\label{tab:results}
\end{table}

Table~\ref{tab:results} shows the total amount of data transfer (GB) for each of the three workloads we have considered. Since queries in the TCP are only mildly overlapping in semantics, the scan cache results in marginal savings when compared to the baseline, while \textit{Bauplan} differential cache translates into up to 30\% savings on S3 reads. To verify the semantic correctness of the cache and the upper bound on savings in the NYC taxi dataset, we compute optimal caching plans by hand and verify that our cache saves as much data as theoretically possible (1.171 GB).

\section{Related work}

\textbf{Pipeline frameworks}. Pipelines as chained transformations became popular in recent Python \cite{airflow} \cite{luigi} and SQL frameworks \cite{dbt}. \textit{Bauplan} is both data-aware in its programming model, and data optimized in the compute layer: existing frameworks leave to the user to glue together code, infrastructure and data cache, often resulting in sub-optimal practices, such as the coupling of physical representations with transformation code. 

\textbf{Differential cache}. While SQL-based caches have been proposed in the database community for OLTP systems on local storage  \cite{10.5555/645922.673462}, the columnar nature of OLAP workloads and the high latency cost of cloud storage sparked most of the recent interest for efficient I/O operations. For example, the OLAP cache for Trino \cite{298597} proposes a file-based, SQL cache for distributed engines -- compared to our proposal, it is not just significantly more complex to integrate, but is neither column-aware nor differential. The \textit{Redshift} predicate cache
\cite{10.1145/3626246.3653395} is more lightweight and fine-grained than Trino's, but still works only on exact filter matching in SQL, as expected from a warehouse-centric design. In contrast, our cache natively supports open table formats in a multi-language lakehouse architecture; as such, it is able to provide a composite view of a data asset, assembled across subsequent scans in real-time. A differential proposal has been recently put together by \textit{MotherDuck} \cite{mduck}: their design is however coupled to SQL and \textit{DuckDB}, and it targets primary storage in a proprietary format.

\section{Conclusion and future work}
In \textit{this} paper we introduced \textit{Bauplan} programming model for data pre-processing in a lakehouse: by making pipelines data-aware, users gain an intuitive semantics without concerns in data management. Recognizing that object storage latency disrupt the  feedback loop, we introduced a new \textit{columnar} and \textit{differential} cache, specifically designed to leverage the strengths of open table formats and power heterogeneous use cases on the lakehouse. Our cache is straightforward to integrate and work across table versions, users and languages (SQL and Python).  Our preliminary benchmarks are encouraging, especially considering that many opportunities for efficiency are yet to be exploited. For example, Iceberg scans do not guarantee row order, making positional joins non-deterministic when a scan adds projections over existing (pre-filtered) tuples. As we continue to collect real-world traces, we look forward to sharing further results with the community and potentially release novel datasets specifically focused on data pre-processing workloads.

\bibliographystyle{IEEEtran}
\bibliography{test}

\end{document}